\documentclass[11pt]{article}
\usepackage{amssymb}
\usepackage{amsmath}
\usepackage[english]{babel}
\usepackage{graphicx}

\usepackage{textcomp}

\usepackage{cite}

\def\g{{\,\rm g}}
\def\eV{\,{\rm eV}}

\def\mean#1{\left< #1 \right>}
\def\({\left(}
\def\){\right)}
\def\cm{{\,\rm cm}}

\def\beq{\begin{equation}}
\def\eeq{\end{equation}}
\def\bea{\begin{eqnarray}}
\def\eea{\end{eqnarray}}

\def\tz{\tilde{z}}

\begin{document}
\title{Primordial black holes with mass $10^{16}-10^{17}$ g and reionization of the Universe}

\author{K.~M. Belotsky$^{1,2}$\thanks{k-belotsky@yandex.ru}, A.~A. Kirillov$^{1}$\thanks{kirillov-aa@yandex.ru}\\
$^1$ National Research Nuclear University MEPhI\\ (Moscow Engineering Physics Institute),\\ Moscow, Russia\\
$^2$ Centre for Cosmoparticle Physics ``Cosmion'',\\ Moscow, Russia\\}

\maketitle

\begin{abstract}
Primordial black holes (PBHs) with mass $10^{16}-10^{17}$ g almost escape constraints from observations so could essentially contribute to dark matter density. 
Hawking evaporation of such PBHs produces with a steady rate $\gamma$- and $e^{\pm}$-radiations in MeV energy range, which can be absorbed by ordinary matter. Simplified estimates show that a small fraction of evaporated energy had to be absorbed by baryonic matter what can turn out to be enough to heat the matter so it is fully ionized at the redshift $z\sim 5\ldots 10$. The result is found to be close to a borderline case where the effect appears, what makes it sensitive to the approximation used. In our approximation, degree of gas ionization reaches 50-100\% by $z\sim 5$ for PBH mass $(3\ldots7)\times 10^{16}$ g with their abundance corresponding to the upper limit.
%
%

\end{abstract}
\section{Introduction}

Modern observations show that the most of baryons are present in intergalactic medium in the form of ionized gas. It was ionized in the period $z\sim 6\ldots 10$, while exact moment and how fast it happens are not known \cite{2001PhR...349..125B}. So far there is no unambiguous understanding of the reasons of reionization \cite{2001PhR...349..125B, 2007MNRAS.382..325B}.  
It is widely supposed that ultraviolet radiation of first stars is responsible for intergalactic medium ionization \cite{2003ApJ...584..621V, 2004MNRAS.350...47S}.
However it is too difficult to get significant ionization fraction under these assumptions \cite{2004MNRAS.350..539R, 2004ApJ...616L..87F}.
Galaxies with low luminosity and active star formation at high $z$ could also contribute significantly to the process of reionization \cite{2006ARA&A..44..415F, 2007ApJ...663...10S}.

Quasars and accreting PBHs are suggested as another possible sources of reionization \cite{2004MNRAS.350..539R, 2004ApJ...613..646D, 2005tmgm.meet..422D, 2007MNRAS.382..325B, 2008ApJ...680..829R}. Under the usual assumptions the ionizing ability of the quasars is 
insufficient to completely ionize the matter at $z\sim 6$ \cite{2001AJ....122.2833F, 2003ApJ...586..693W, 2005MNRAS.356..596M}. The quasar spectrum analysis at $z > 6$ indicates that intergalactic medium had been significantly ionized before the quasars could make it \cite{2007AJ....134.2435W}. However, quasars could play its role in early reionization under certain conditions (for example, \cite{2004ApJ...604..484M}). In turn, the quasar formation might be associated with the most massive PBH clusters \cite{2000hep.ph....5271R, 2001JETP...92..921R, 2005APh....23..265K, 2005GrCo...11...99D}.

Annihilating dark matter has been considered also as a possible source of reionzation \cite{2010PhRvD..81l3510N, 2009JCAP...10..009C} (and references therein).

In this work we consider the possibility of reionization of the Universe with the help of PBHs, which mass is around $M_{17}=10^{17}$ g. PBHs of this mass range escape constraints from observations on gamma-radiation data and lensing effects 
\cite{2010PhRvD..81j4019C} 
so can provide all dark matter. 
PBHs with mass $> 10^{17}$ g are attempted to be constrained on the base of a tidal capture of PBH by neutron stars \cite{2014JCAP...06..026P}, however it met counterarguments  \cite{2014arXiv1403.7098C, 2014arXiv1409.0469D}.

The PBHs with mass $\sim 10^{17}$ have more attractive features: with the help of them one could explain positron line from Galactic center \cite{2006A&A...450.1013W} due to effects of accretion \cite{2006ApJ...641..293T} or Hawking evaporation \cite{2008PhLB..670..174B}.
Here we probe reionization possibility 
on the base of Hawking evaporation effect. Earlier \cite{2015PAN............} we discussed similar possibility for a cluster of PBHs with a power-law mass spectrum, which was predicted in \cite{2000hep.ph....5271R, 2001JETP...92..921R, 2005APh....23..265K} and used in a treatment of unidentified cosmic gamma-ray sources \cite{2011GrCo...17...27B, 2011APh....35...28B}. The result of \cite{2015PAN............} did not allow to make definite conclusion, but it was seen that existing observational constraints on PBH mass spectrum and the freedom in its theoretical predictions leave potential for possible solution of reionization problem.

We assume that PBHs have narrow mass distribution ($\delta$-functional-like). The mechanism of their production is not discussed, different ones are reviewed in \cite{2005astro.ph.11743C, 2010RAA....10..495K, 2014arXiv1410.0203B, 2010PhRvD..81j4019C, 2007arXiv0709.0070D}.


\section{General approach}

In our estimations we take the mass range $10^{16}\g<M<10^{17}\g$ and abundance according to the upper limit given in \cite{2010PhRvD..81j4019C}, which we present in the form
\begin{equation}
\Omega_{\rm PBH}=\left\{
\begin{array}{l}
0.25,\text{ if $M>M_{\rm peak}$} \\
0.25\left(\frac{M}{M_{\rm peak}}\right)^{3.36},\text{ if $M<M_{\rm peak}$,}
\end{array}
\right.
\label{OmegaPBH}
\end{equation}
where $M_{\rm peak}=0.78\times 10^{17}$ g.

The evaporation temperature for such PBH is $T_{\rm ev}\approx 0.106\frac{M_{17}}{M}$ MeV, the mean energy of evaporating particles is 
$\approx 6T_{\rm ev}$ for photons and $\approx 4T_{\rm ev}$ for electrons and neutrinos. Emitted $e^{\pm}$ energy is to be corrected, when $T_{\rm ev}\lesssim m_e$.

In reionization scenarios with first stars and accreting black holes, ionizing radiation is basically a short-range ultraviolet, which arises locally in the regions of strong inhomogeneities formation. It leads to a complicated picture of inhomogeneous Universe ionization. 
In the case of evaporated PBHs, it is not so, and ionization (effects of Hawking radiation interaction with matter) can be supposed to proceed homogeneously over all volume.

Two simplified ways to estimate reionization possibility can be suggested. Ionizing particle induces one ionization in usual matter per each 20--40~eV of the lost energy \cite{Blum}. So one could consider, ignoring details of dissipation processes in matter, that 20 eV of evaporated radiation energy absorbed in matter produces irreversibly one ionization act.
Full ionization of baryonic matter in the Universe can be supposed to happen when the total absorbed energy reaches 20 eV per each atom. So, in units of critical density of Universe ($\rho_c$) it is
\begin{gather}
\label{OmegaCr}
\Omega_{\rm abs}\gtrsim \Omega_{\rm ion}\\
\Omega_{\rm ion}=\frac{20 \eV}{m_p}\Omega_B\approx 10^{-9}.
\label{OmegaIon}
\end{gather}

Certainly, energy release 20 eV per each atom is more than enough for ionization of gas, provided thermodynamic equilibrium is in absence of strong cooling.
But in our case the process of heating from PBHs is extended in time and goes 
against permanent cooling due to expansion of the Universe. 
%
%
Similarly, we apply pure temperature arguments in the second way. It is considered, that the energy release induced by particle from PBH is quickly transformed by thermalization processes (including ionization and recombination) into heat, and ionization fraction of matter is defined by Saha formula. 

In the both ways, one should take into account that not all initial energy of evaporated particle can be deposited in baryonic matter. This process takes finite time, so it may not be completed, and also part of energy can be lost due to red shift or other processes.

\section{Estimations}

Let us denote the rates of energy evaporation and absorption by matter in units of critical density as $\dot{\Omega}_{\rm ev}$ and $\dot{\Omega}_{\rm abs}$. To estimate the first one, one approximates single PBH evaporation rate as
\begin{equation}
	\dot{M}= \frac{1}{3}\(\frac{M_U}{M}\)^2\frac{M_U}{t_U},
	\label{dotM}
\end{equation} 
where $M_U\approx 0.5\times 10^{15}$ g is the mass of PBH which would evaporate fully by present time, $t_U\approx 14$ Gyr is the Universe age.
Eq.~\eqref{dotM} must be multiplied by the ratio of effective numbers of evaporated particle species for $M_U$ and $M$, $g_{\rm tot}(M)/g_{\rm tot}(M_U)\sim 1$. But we will effectively take it into account normalizing the fractions of evaporated particle species by $g_{\rm tot}(M_U)$. 
The rate $\dot{M}$ is independent of time in given approximation what is good when $M\gg M_U$. The same is applied to the total evaporation rate:
\begin{equation}
\dot{\Omega}_{\rm ev}=\frac{\dot{M}}{M}\Omega_{\rm PBH}(M)
= \frac{1}{3}\(\frac{M_U}{M}\)^3\frac{\Omega_{\rm PBH}(M)}{t_U}.
\end{equation}
Total energy to be released in evaporation $\Omega_{\rm ev}=\dot{\Omega}_{\rm ev}t_U\approx 10^{-8}\(\frac{M_{17}}{M}\)^{3}\frac{\Omega_{\rm PBH}}{0.25}$ has some 'reserve' over $\Omega_{\rm ion}$ what makes mechanisms of settling this energy in matter of special importance.

In temperature interval of question, PBH emits gravitons with freedom degrees weight $g_G=2\cdot0.007$, photons with $g_{\gamma}=2\cdot0.06$, three sorts of neutrinos with $g_{\nu\bar{\nu}}=6\cdot0.147$ and electrons and positrons with $g_{e^{\pm}}=4\cdot0.142\cdot \hat{g}_e(m_e/T)$, where $\hat{g}_e(m_e/T)$ takes into account partial suppression of massive electron production. We shall take it as
\begin{equation}
\hat{g}_e(M)=\(1+0.40\frac{M}{M_{17}}\)^{27.6}\exp\(-10.9\frac{M}{M_{17}}\)
\label{ge}
\end{equation}
for the considered $M$.
$g_{\rm tot}(M_U)\approx 1.6$ is defined as a sum of all the weights at  $\hat{g}_e(m_e/T)=1$.
Fractions of $e^{\pm}$ and photons in evaporation flux are respectively
\begin{equation}
\kappa_{e^{\pm}}=g_{e^{\pm}}/g_{\rm tot}(M_U)\approx 0.36\hat{g}_e(M)
,\;\;\;\;     \kappa_{\gamma}=g_{\gamma}/g_{\rm tot}(M_U)\approx 0.08.
\end{equation}

\begin{figure}
	\centering
	\includegraphics[scale=1]{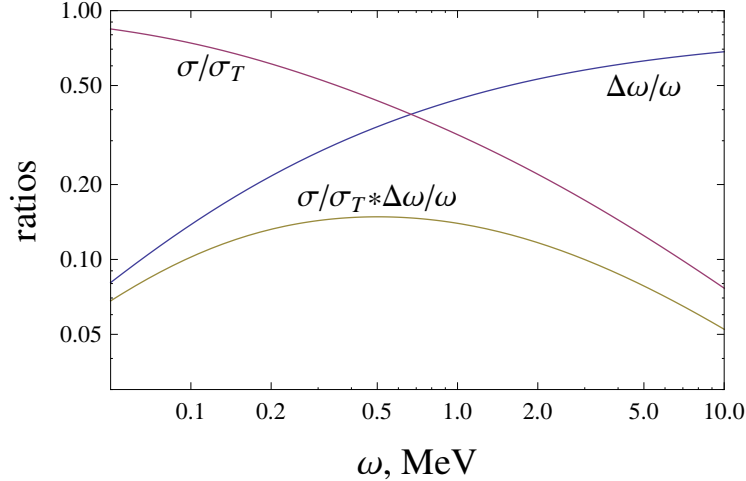}
	\caption{The ratio $s=\sigma/\sigma_T$, the mean relative energy transfer $q=\Delta\omega/\omega$ and their product $s\cdot q$ for Compton process is shown depending on photon energy.}
	\label{ratio}
\end{figure}

\bigskip
{\it \textbf{Photons}}
\medskip

The main interaction process of photon from evaporation in energy range of question ($\omega\sim 0.5\dots 5$ MeV) is the Compton scattering. In this energy range the total cross section (Klein-Nishina one), $\sigma=s(\omega)\sigma_T$, deviates from Thomson one by factor $s(\omega)\sim 0.1\ldots 0.5$, which falls with photon energy $\omega$ growth (see Fig.~(\ref{ratio})). Mean relative energy transfer in one scattering, $q=\Delta\omega/\omega$, is also energy dependent but grows with $\omega$, as shown on the same figure. For calculations we shall put $sq=\text{const}=0.1$ (see Fig.~(\ref{ratio})). Then respective energy loss time scale is given by $\tau_C=(n_H\sigma_T (s q) c)^{-1} = t_U\tz_C^{3/2}\tz^{-3}$,
where $n_H=x_H\cdot n_B^{\rm mod}\tz^3$ is the total number density of hydrogen with $n_B^{\rm mod}=2.5\times 10^{-7}\cm^{-3}$ and $x_H=0.76$ being the modern baryon number density and hydrogen fraction in it, $\tz_C\approx 340$. Here and throughout we accept denotation $\tz_{(i)}\equiv z_{(i)}+1$, and that MD stage takes place only, helium component is not considered.

Let the portion of energy $\delta \Omega_{\gamma}(t_0)=\kappa_{\gamma}\dot\Omega_{\rm ev}dt_0$ per unit mass of matter in the Universe be released by PBHs at the moment $t_0$ in the form of photons. Decrease rate of this portion is defined by Compton scattering and red shift, so one writes
\begin{equation}
\frac{d\,\delta \Omega_{\gamma}(t)}{dt}=-\frac{\delta \Omega_{\gamma}(t)}{\tau_C}-H\,\delta \Omega_{\gamma}(t),
\label{ddelta}
\end{equation}
where 
$H=2/3\,t_U^{-1}\tz^{3/2}$ is the Hubble parameter. Note, the first term in the r.s. of Eq.~\eqref{ddelta} characterizes the rate of energy absorption by matter. 
Solving of Eq.~\eqref{ddelta} in variable $\tz=(t_U/t)^{2/3}$ gives
\begin{equation}
\delta\Omega_{\gamma}(z_0,z)=\frac{\tz}{\tz_0}\exp\(-\frac{\tz_0^{3/2}-\tz^{3/2}}{\tz_C^{3/2}}\)\cdot \kappa_{\gamma}\dot{\Omega}_{\rm ev}dt_0.
\label{delta}
\end{equation}

Integrating of Eq.~\eqref{delta} over $t_0$ in the interval preceding to $t$ ($t_i(z_i)<t_0(z_0)<t(z)$) gives the total energy of photons from evaporation present at the moment $t(z)$, $\Omega_{\gamma}(z)$. 
%
%
The values $\dot{\Omega}_{\rm abs}^{(\gamma)}(z)\equiv \Omega_{\gamma}(z)/\tau_C $ and $\Omega_{\rm abs}^{(\gamma)}(z)=\int_{t_i}^{t(z)} \dot{\Omega}_{\rm abs}^{(\gamma)}(t'(z'))dt'$ give respectively the total energy absorption rate depending on $z$ and the total photon energy absorbed by the moment $z$. Explicitly they are
\begin{gather}
\label{ROmega}
\dot{\Omega}_{\rm abs}^{(\gamma)}(z)=
\kappa_{\gamma}\dot{\Omega}_{\rm ev}\cdot f_{\rm abs}^{(\gamma)}(z)\\
f_{\rm abs}^{(\gamma)}(z)=\int_{z}^{z_i}\frac{\tz}{\tz_0}\exp\(-\frac{\tz_0^{3/2}-\tz^{3/2}}{\tz_C^{3/2}}\)\frac{3t_Ud\tz_0}{2\tz_0^{5/2}\tau_C}= \nonumber\\
\label{fabs}
=\frac{\tz^4}{\tz_C^4}\exp\(\frac{\tz^{3/2}}{\tz_C^{3/2}}\)\left[\Gamma\(-\frac{5}{3},\frac{\tz^{3/2}}{\tz_C^{3/2}}\)-\Gamma\(-\frac{5}{3},\frac{\tz_i^{3/2}}{\tz_C^{3/2}}\) \right],\\
\Omega_{\rm abs}^{(\gamma)}(z)=\kappa_{\gamma}\dot{\Omega}_{\rm ev}t_U \cdot \int_{\tz}^{\tz_i} f_{\rm abs}^{(\gamma)}(z')\frac{3d\tz'}{2\tz'^{5/2}}.
\label{Omega}
\end{gather}
Note that $f_{\rm abs}^{(\gamma)}$ tends to 1 while $z_C/(z_i-z)\rightarrow 0$.

\begin{figure}
	\centering
	\includegraphics[scale=1]{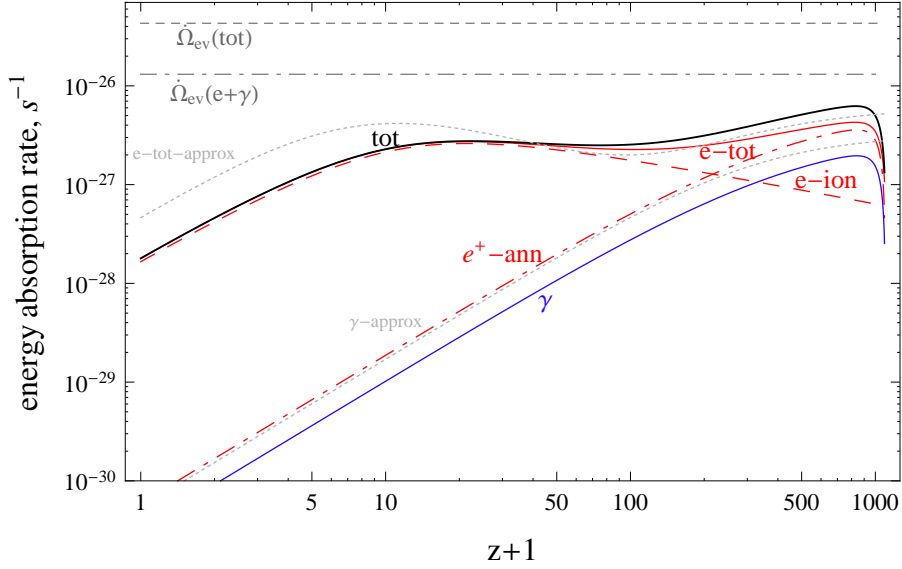}
	\caption{ Energy absorption rates for all processes considered: $\dot{\Omega}_{\rm abs}^{(e{\rm-ion})}$, $\dot{\Omega}_{\rm abs}^{(e{\rm-ann})}$ and their sum, $\dot{\Omega}_{\rm abs}^{(\gamma)}$ and the total rate, for $M=5\times 10^{16}$ g. $\dot{\Omega}_{\rm abs}^{(e{\rm-ion})}+\dot{\Omega}_{\rm abs}^{(e{\rm-ann})}$ and $\dot{\Omega}_{\rm abs}^{(\gamma)}$, obtained with Eqs.~\eqref{apprG}--\eqref{apprAnn}, are also shown. $\dot{\Omega}_{\rm ev}$ is shown to illustrate total evaporation rate in all species and $e^{\pm}+\gamma$ only.}
	\label{rates}
\end{figure}

\begin{figure}
	\centering
	\includegraphics[scale=1]{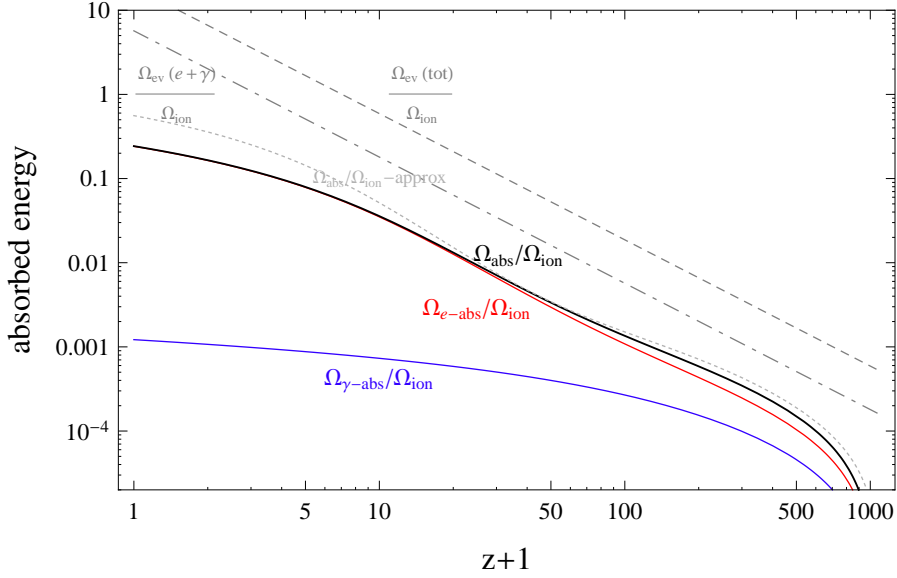}
	\caption{Evaporated electron and photon energy absorbed by baryonic matter in units of $\Omega_{\rm ion}$: $\Omega_{\rm abs}^{(e)}$, $\Omega_{\rm abs}^{(\gamma)}$ and their sum, for $M=5\times 10^{16}$ g. Total absorbed energy, obtained with Eqs.~\eqref{apprG}--\eqref{apprAnn}, is also shown. $\Omega_{\rm ev}$ is shown, illustrating the total energy being emitted by PBHs.}
	\label{Omegas}
\end{figure}

The value $\dot{\Omega}_{\rm abs}^{(\gamma)}(z)$ is shown on Fig.~(\ref{rates}) in comparison with $\dot{\Omega}_{\rm ev}$. The ratio $\frac{\Omega_{\rm abs}^{(\gamma)}(z)}{\Omega_{\rm ion}}$ is shown on the Fig.~(\ref{Omegas}). Initial moment is formally taken to be $z_i=1100$.

\bigskip
{\it \textbf{Electrons and positrons}}
\medskip

Electrons and positrons from evaporation of PBH should experience energy losses due to scattering off cosmic microwave background (CMB) photons, ionization and red shift. Effects of interaction with low-density plasma is not considered here. 

Energy losses on CMB in ultra-relativistic limit are given by \cite{berezinsky1990astrophysics}
\begin{equation}
\left.\frac{dE}{dt}\right|_{\rm CMB}=-\beta E^2,
\label{cmb}
\end{equation}
where $\beta= 
\omega_2^{-1} t_U^{-1}\tz^4$ is defined by CMB energy density, $\omega_2\approx 90$ MeV. Note, each scattering leads to energy transfer as small as $\sim (E/m)^2$ of primary CMB photon energy. Half energy loss time is given by $\tau_{\rm CMB}=(\beta E)^{-1}$. 

Ionization losses rate is approximated by its minimal value for hydrogen $dE/dx=\text{const} \approx 4$ MeV g$^{-1}$cm$^2$ until the ionizing particle stops, so
\begin{equation}
\left.\frac{dE}{dt}\right|_{\rm ion}=-\frac{dE}{dx}\frac{dx}{dt}=-r_{\rm ion}\eta(E) 
\label{ion}
\end{equation}
where $r_{\rm ion}=\frac{dE}{dx}cn_Hm_p 
\approx \omega_1 t_U^{-1}\tz^3$ with $\omega_1=0.016$ MeV, $\eta$ is the step function. 

\begin{figure}
	\centering
	\includegraphics[scale=1]{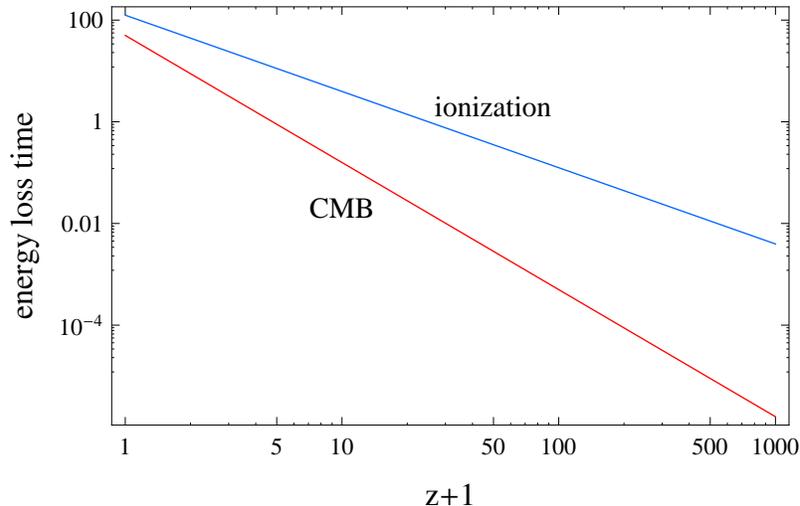}
	\caption{Energy loss time of electron with $E=2$ MeV is shown in units of cosmological time.}
	\label{e-length}
\end{figure}

The values $\tau_{\rm CMB}$, $\tau_{\rm ion}=E/(dE/dt)_{\rm ion}$ at $E=2$ MeV in units of cosmological time $t=t_U\tz^{-3/2}$ are shown on the Fig.~(\ref{e-length}). As seen, 
the losses on CMB dominate until the late period (small $z$), where both CMB scattering and ionization losses rates become slower than that of Universe expansion. 
Note, that only relatively small ionization losses are supposed to provide energy transfer to the baryonic matter.

For PBH mass $\sim 10^{17}$ g, electrons are emitted in sub-relativistic regime. We accept approximation referring everywhere to $E$ as kinetic energy, and considering full electron energy at evaporation to be $4T\hat{h}_e(M)+m_e$. (Nonetheless, we treat energy losses ultra-relativistically; the impact of this approximation on the result is discussed below.) Here $\hat{h}_e(M)$ takes into account a decrease of evaporated electron kinetic energy with respect to $4T_{\rm ev}$  when $T_{\rm ev} \lesssim m_e$, and is taken for the considered range of $M$ in the form
\begin{equation}
\hat{h}_e(M)=\exp\(-1.45\frac{M}{M_{17}}-\frac{M^2}{M_{17}^2}\).
\label{he}
\end{equation} 
The functions $\hat{h}_e(M)$ and $\hat{g}_e(M)$ were chosen to roughly fit the spectrum of evaporated electrons with properties such as: absorption black hole cross section changes in $27/2$ times with the growth of energy from $E\ll T_{\rm ev}$ to $E\gg T_{\rm ev}$, and at $E=m_e$ absorption probability is about 0.5 of its relativistic value \cite{1990PhRvD..41.3052M}.

To write down equation analogous to Eq.~\eqref{ddelta} for energy portion of electrons, 
one needs to make replacement $E\rightarrow \frac{E_0}{\delta\Omega_e(t_0)}\delta\Omega_e(t)$ in respective equation for the single electron losses, where $E_0\approx 4 T\hat{h}_e(M)$ and $\delta\Omega_e(t_0)=\kappa_e\dot{\Omega}_{\rm ev}dt_0\frac{E_0}{E_0+m_e}$ are the initial kinetic energy of one evaporated electron and all of them (per unit mass $\rho_cV$). So
\begin{gather}
\frac{d\,\delta \Omega_e(t)}{dt}= 
-\frac{r_{\rm ion}\delta\Omega_e(t_0)}{E_0}\eta\({\delta\Omega_e(t)}\)-\frac{\beta E_0}{\delta\Omega_e(t_0)}\delta\Omega_e^2(t)-H\,\delta \Omega_e(t).
\label{difOmega}
\end{gather}

Exact solution of this equation is cumbersome. But we need to know only first term of r.s. of Eq.\eqref{difOmega}, which is already fixed and only the moment $t=t_s(z_s)$, when $\delta\Omega_e(t)=0$ (the particle stops), is to be determined. It can be found from equation
\begin{equation}
\delta\Omega_e(t_0)=\delta\Omega_e^{(1)}(t)+\delta\Omega_e^{(2,3)}(t),
\label{Omega0}
\end{equation} 
where 
\begin{gather}
\delta\Omega_e^{(1)}(t(z))=\delta\Omega_e(t_0(z_0))\(1-\frac{2\omega_1}{3E_0}\tz_0^{3/2}\(1-\frac{\tz^{3/2}}{\tz_0^{3/2}}\)\) \\
\delta\Omega_e^{(2,3)}(t(z))=\frac{\tz}{\tz_0} \frac{\delta\Omega_e(t_0(z_0))}{1+\frac{2E_0}{7\omega_2}\tz_0^{5/2}\(1-\frac{\tz^{7/2}}{\tz_0^{7/2}}\)}
\label{Omega123}
\end{gather}
are the solutions of Eq.~\eqref{difOmega}, when only first or second and third terms are present in r.s. Since the particle stops rapidly over the most period of question (see Fig.~(\ref{e-length})), we can take approximation $\tz_s= \tz_0(1-\zeta_0)$ with $\zeta_0 \ll 1$, or if we fix $z=z_s$ then maximal $\tz_0$ is $\tz_{\rm 0max}= \tz(1+\zeta)$ with $\zeta \ll 1$. For further calculation, one finds $\zeta$
from Eq.~\eqref{Omega0}
\begin{equation}
\zeta=\frac{-1+\sqrt{ 1+\frac{E_0}{\omega_1 \tz^{3/2}} \(1+\frac{E_0}{\omega_2}\tz^{5/2}\) } }{2\(1+\frac{E_0}{\omega_2}\tz^{5/2}\)}.
\end{equation}
At $z\sim 10$ it gives result with 30\% accuracy, at higher $z$ error quickly tends to zero.

Step-function in Eq.~\eqref{difOmega} is reduced to $\eta(z_{\rm 0max}-z_0)$. Thus, trivial integrating of the first term in r.s. of Eq.~\eqref{difOmega} over $z_0(t_0)$ in the period $z<z_0<z_{\rm 0max}$ gives the total ionization rate (energy absorption)
\begin{equation}
\dot{\Omega}_{\rm abs}^{(e{\rm-ion})}(z)=\kappa_e\dot{\Omega}_{\rm ev}\frac{\omega_1}{E_0+m_e}\frac{3\zeta(z)\tz^{3/2}}{2+3\zeta(z)}.
\label{ROmegaIon}
\end{equation}
Note, that the last fraction in Eq.~\eqref{ROmegaIon} gives good approximation also when $\zeta>1$. 
The value $\dot{\Omega}_{\rm abs}^{(e{\rm-ion})}(z)$ 
is shown on the Fig.~(\ref{rates}). Behaviour of  $\dot{\Omega}_{\rm abs}^{(e{\rm-ion})}(z)$ with decreasing $z$ can be traced with Fig.~(\ref{e-length}): at high $z$ the rate grows because of relative growth of ionization losses on the background of dominating losses on CMB, and at low $z$ the energy is lost due to red shift.
Total energy of electrons absorbed by the moment $z$ is given by $\Omega_{\rm abs}^{(e{\rm-ion})}(z)=\int_{\tz}^{\tz_i}\dot{\Omega}_{\rm abs}^{(e{\rm-ion})}(\tz')\frac{3t_Ud\tz'}{2\tz'^{5/2}}$.

One more contribution into heat of matter should be given by the annihilation of the stopped positrons from PBHs. It is not difficult to estimate it. Each from $\frac{1}{2}\frac{\kappa_e\dot{\Omega}_{\rm ev}dt_0}{E_0+m_e}(\rho_cV)$ positrons, evaporated within interval $dt_0$, produces $2m_e$ energy in form of gamma, which absorption during their propagation is described by Eqs.~\eqref{ROmega}--\eqref{Omega}. So, for energy absorption rate and total absorbed energy by this mechanism one has
\begin{gather}
\dot{\Omega}_{\rm abs}^{(e{\rm-ann})}(z)=\kappa_e\dot{\Omega}_{\rm ev}\frac{m_e}{E_0+m_e}f_{\rm abs}^{(\gamma)}(z),\\
\Omega_{\rm abs}^{(e{\rm-ann})}(z)=\int_{\tz}^{\tz_i}\dot{\Omega}_{\rm abs}^{(e{\rm-ann})}(z')\frac{3t_Ud\tz'}{2\tz^{5/2}}.
\end{gather}

Total absorption rate is given by the sum $\dot{\Omega}_{\rm abs}=\dot{\Omega}_{\rm abs}^{(\gamma)}+\dot{\Omega}_{\rm abs}^{(e\text{-ion})}+\dot{\Omega}_{\rm abs}^{(e\text{-ann})}$. All the energy absorption rates are shown on the Fig.~\ref{rates}, total absorbed energies in units $\Omega_{\rm ion}$ are on the Fig.~\ref{Omegas}, for each mechanism and their sums. PBH mass is taken to be $M=5\times 10^{16}$ g and abundance is from  Eq.~\eqref{OmegaPBH}.
As seen, the total absorbed energy reaches 5-10\% from $\Omega_{\rm ion}$ at $z\sim 10$, what does not allow to make conclusion about reionization of Universe with the first criterion Eq.~\eqref{OmegaCr}.

\bigskip
{\it \textbf{Order-of-magnitude check}}
\medskip

Absorption rate $\dot{\Omega}_{\rm abs}^{(i)}$ due to energy loss process "$i$" can be roughly estimated as the 
evaporation rate of the respective particles multiplied by the fraction of the energy loss rate due to $i$-process in the total relevant energy losses rate.
Each rate can be roughly estimated as respective inverse characteristic time. So for the rates of question one has
\begin{gather}
\label{apprG}
\dot{\Omega}_{\rm abs}^{(\gamma)}\sim 
\kappa_{\gamma}\dot{\Omega}_{\rm ev}\frac{\tau_C^{-1}}{\tau_C^{-1}+t^{-1}}= \kappa_{\gamma}\dot{\Omega}_{\rm ev}\frac{\tz^{3/2}}{\tz^{3/2}+\tz_C^{3/2}},\\
\label{apprIon}
\dot{\Omega}_{\rm abs}^{(e{\rm-ion})}\sim \kappa_e\dot{\Omega}_{\rm ev}\frac{\tau_{\rm ion}^{-1}}{\tau_{\rm ion}^{-1}+\tau_{\rm CMB}^{-1}+t^{-1}}=\kappa_e\dot{\Omega}_{\rm ev}\frac{\frac{\omega_1}{E_0}\tz^{3/2}}{\frac{\omega_1}{E_0}\tz^{3/2}+\frac{E_0}{\omega_2}\tz^{5/2}+1},\\
\label{apprAnn}
\dot{\Omega}_{\rm abs}^{(e{\rm-ann})}\sim \kappa_e\dot{\Omega}_{\rm ev}\frac{m_e}{E_0+m_e}\frac{\tz^{3/2}}{\tz^{3/2}+\tz_C^{3/2}}.
\end{gather}
Here $t^{-1}$ characterizes red shift rate. $\dot{\Omega}_{\rm abs}^{(\gamma)}$ and $\dot{\Omega}_{\rm abs}^{(e{\rm-ion})}+\dot{\Omega}_{\rm abs}^{(e{\rm-ann})}$ given by Eqs.~\eqref{apprG}--\eqref{apprAnn} are shown on the Fig.~(\ref{rates}). Total absorbed energy, obtained from Eqs.~\eqref{apprG}--\eqref{apprAnn}, is shown on the Fig.~(\ref{Omegas}). As seen, at given $M$ different approximations keep within factor 3.

\bigskip
{\it \textbf{Termodynamical consideration}}
\medskip

Let us estimate the temperature of baryonic matter (hydrogen). One takes formally the first law of thermodynamics, $dQ=\delta A+dU$, for arbitrary amount of matter $n_H V$ (here, as previously, index $H$ relates to both of atomic hydrogen and protons). One has the electron fraction $x_e=n_e/n_H$, full number density of particles in plasma $n_m=n_H(1+x_e)$, the pressure $p=n_mT$. The temperatures of atomic hydrogen and ion-electron component are assumed to be equal. 

Expansion of the Universe is treated as a work of gas: $\delta A=pdV=n_mT\, 3HVdt$. Inner energy gain of gas is $dU=\frac{3}{2}d(pV)=(p=n_mT,\; n_mV\approx\text{const})=\frac{3}{2}n_mVdT$. The heat gain is $dQ=\dot{\Omega}_{\rm abs}\rho_cVdt-\mean{\Delta E\sigma v}_{m\gamma}n_{\gamma}n_eVdt$, where the second term takes into account CMB-matter(electrons) energy exchange, $\mean{\Delta E\sigma v}_{m\gamma}n_{\gamma}=\frac{4\pi^2}{15}T_{\gamma}^4\sigma_T\frac{T-T_{\gamma}}{m_e}$. Substituting it all in the first law of thermodynamics and making simple transformations one gets
\begin{equation}
\frac{dT}{dt}=\frac{2\dot{\Omega}_{\rm abs}m_p}{3x_H\Omega_B(1+x_e)}-\frac{8\pi^2}{45}T_{\gamma}^4\sigma_T\frac{x_e}{1+x_e}\frac{T-T_{\gamma}}{m_e}-2HT.
\label{dT}
\end{equation}

The value $x_e$ is defined from Saha formula
\begin{gather}
\frac{x_e^2}{1+x_e}=\frac{1}{n_H}\(\frac{m_eT}{2\pi}\)^{3/2}\exp\(-\frac{Rg}{T}\)\approx\\ \nonumber 
\approx\frac{1.6\times 10^{28}T^{3/2}}{\tz^3}\exp\(-\frac{13.6}{T}\),
\label{Saha}
\end{gather}
where $T$ is in eV.

Basically, first and third terms in r.s. of Eq.~\eqref{dT} are important. If to suppose $\dot\Omega_{\rm abs}=\text{const}$ and to neglect by $x_e$ in the first term, then solution of Eq.~\eqref{dT} without second term would have simple view
\begin{equation}
T(t)=\frac{2\dot{\Omega}_{\rm abs}m_p}{7x_H\Omega_B}\cdot t\(1-\(\frac{t_0}{t}\)^{7/3}\)+T_0\(\frac{t_0}{t}\)^{4/3},
\label{T}
\end{equation}
where $T_0=T(t_0)$ is the initial value. Thus, one would obtain inevitably linear growth of temperature starting from some moment. 
If $\dot{\Omega}_{\rm abs}\sim 10^{-27}$ s$^{-1}$ then the matter is heated by $z\sim 10$ upto temperature to be ionized, what is close to the situation considered (see Fig.~(\ref{rates})).

\begin{figure}
	\centering
	\includegraphics[scale=1]{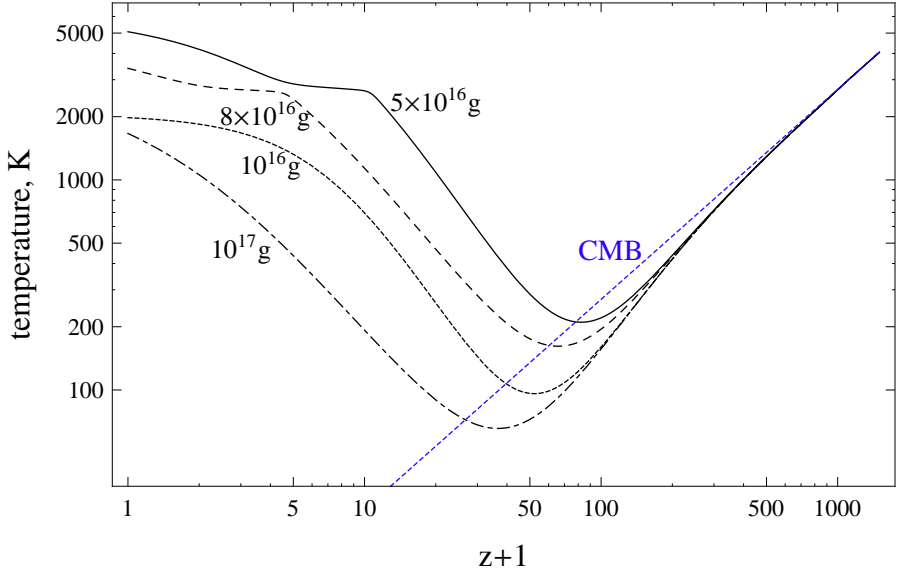}
	\caption{Temperatures of hydrogen at different PBH masses and CMB.}
	\label{Tfig}
\end{figure}

\begin{figure}
	\centering
	\includegraphics[scale=1]{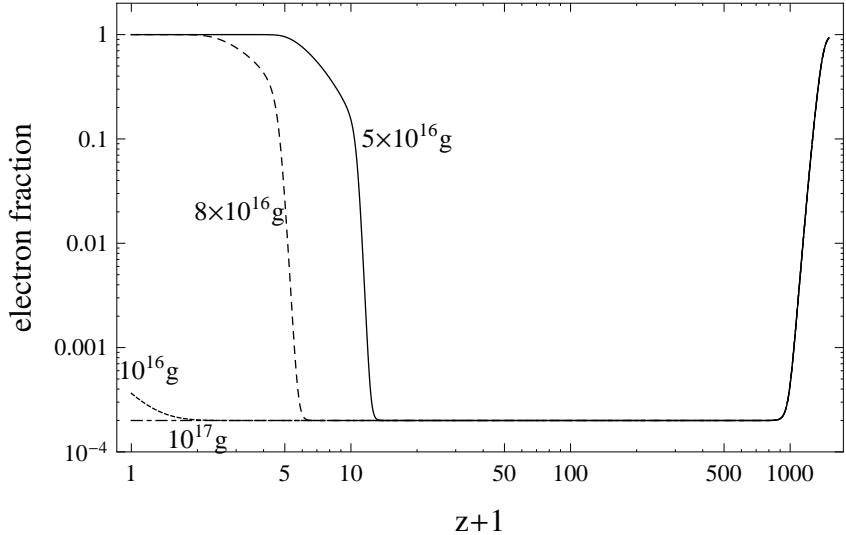}
	\caption{Electron fraction $x_e$ in dependence of the redshift.}
	\label{xe}
\end{figure}

Exact solution of Eq.~\eqref{dT} for matter temperature and electron fraction $x_e$ are shown on the Fig.~(\ref{Tfig}) and (\ref{xe}). The choice of initial conditions has almost no influence. E.g., varying $z_i$ in interval 100--2000 does not virtually affect the result. The main effect sits at $z\sim$~50--100 when expansion rate falls lower than that of heat gain. Minimal electron fraction was formally fixed to be 
$x_e=2\times 10^{-4}$, in accordance with its frozen magnitude after CMB-matter decoupling. 

The used approximation, based on Saha formula, gives that neutral gas-plasma transition happens at temperature similar to recombination one (only logarithmic correction due to difference in number densities). Existing rough estimation of the intergalactic gas temperature gives value at the level $\sim 10^4$ K for $z\approx 4$ \cite{2000MNRAS.318..817S}, what agrees tolerably with our result. Accounting for other sources of the heat can improve agreement.

\begin{figure}
	\centering
	\includegraphics[scale=1]{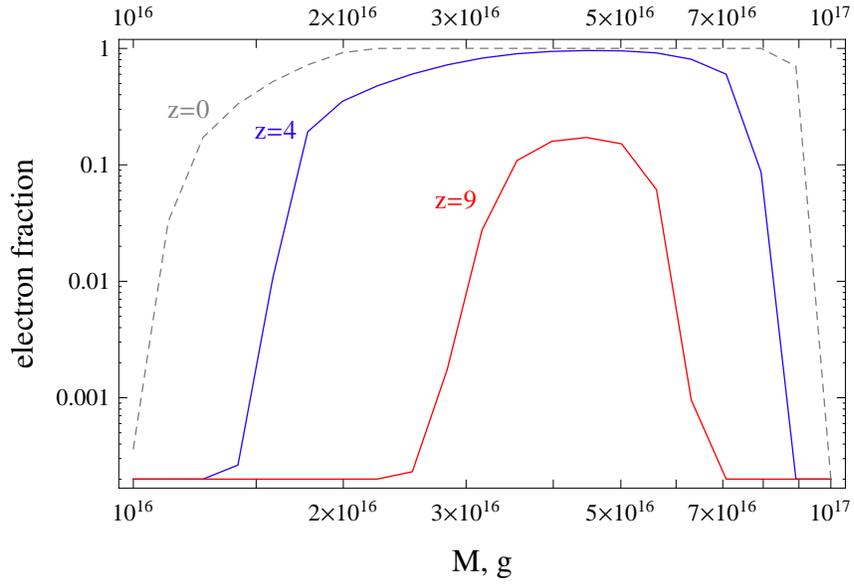}
	\caption{Electron fraction $x_e$ in dependence on PBH mass.}
	\label{xeM}
\end{figure}

At the Fig.~(\ref{xeM}) $x_e$ as function of PBH mass $M$ is shown for $\tz=10$, 5 and 1. $\Omega_{\rm PBH}$ corresponds to the upper limit Eq.~\eqref{OmegaPBH}. As seen, PBH with masses in the range $3\times 10^{16}\dots 7\times 10^{16}$ g could provide reionization of the Universe.

Note, that the result turns out to be sensitive to the estimation of $\dot{\Omega}_{\rm abs}^{(e\text{-ion})}$ which plays the main role in the heating and eventual ionizing of the matter. So, accuracy of all approximations applied becomes important, including that of suppression factors of $e^{\pm}$-yield (Eq.~\eqref{ge}), which suppresses the effect, and their kinetic energy (Eq.~\eqref{he}), which increases the effect due to suppression of competing energy losses (on CMB and red shift), which in turn have stronger energy-dependences (see Eq.~\eqref{difOmega}). 

\begin{figure}
	\centering
	\includegraphics[scale=1]{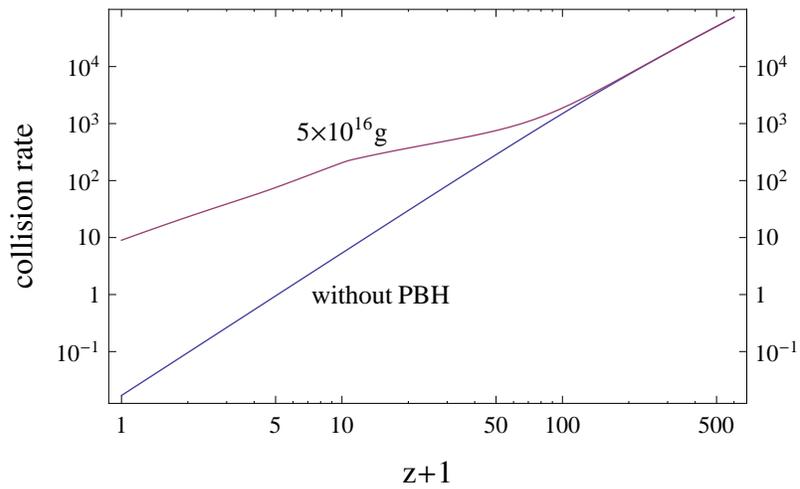}
	\caption{Collision rate of hydrogen atoms in units of cosmological time $t$ as function of the redshift for the cases when PBH mass is $5\times 10^{16}$ g, and there are no PBHs.}
	\label{collrate}
\end{figure}

One of the conditions of thermodynamic equilibrium, under assumption of which the present result is obtained, is a high collision rate of the gas particles (hydrogen atoms,...) as compared to, say, expansion rate. The Fig.~\ref{collrate} shows the number of collisions of hydrogen atoms happening for cosmological time ($t=t_U/\tz^{3/2}$). As seen, this rate could be supposed to be high. If even so, the real process of temperature change and, respectively, electron fraction should have some delay as compared to what shown on the Fig.~(\ref{Tfig}) and (\ref{xe}), because of deviation from perfect equilibrium. 

For large PBH mass ($\sim 10^{17}$ g), evaporated electrons become non-relativistic. We have checked out that ionization losses rate grows relatively to those due to scattering on CMB and red shift while the energy decreases (PBH mass increases). So the used approximation does not overestimate effect. But our check has shown that at $M\sim 10^{17}$ suppression of electrons (Eqs.\eqref{ge},\eqref{he}) becomes so strong that the high-mass slump on the Fig.\ref{xeM} (and, respectively, upper value of the obtained PBH mass interval) does not virtually change even if all electrons would be absorbed immediately.

\section{Conclusion and discussion}

In this paper we considered energetic effects in the baryonic matter induced by evaporation of PBHs of mass ranging $10^{16}\dots 10^{17}$ g. Energetic losses on ionization of evaporated $e^{\pm}$, on Compton scattering of evaporated $\gamma$ and of $\gamma$ from annihilation of evaporated $e^+$ are assumed to go thoroughly into heat of matter (treated as absorbed energy). Ionization losses are found to be the main of them. The energy lost due to ionizations is suppressed, in its turn, by $e^{\pm}$-scattering off CMB photons and, in the later (main in time scale) period, by red shift. Nonetheless, temperature arguments show that ionization degree of matter reaches $\sim 50-100$\% at $z\sim 5$ for PBH mass in interval $(3\ldots7)\times 10^{16}$ g.

The result is found to be close to a borderline case where the effect is or not, what makes it sensitive to the approximations used. We used a  set of simplifying approximations: energetic spectra of evaporated particles are replaced by $\delta$-functions, suppression factors for $e^{\pm}$ yield and energy are taken in the form Eq.~\eqref{ge} and Eq.~\eqref{he}, ionization losses rate is taken to be equal to its minimal value, the losses of $e^{\pm}$ on CMB and red shift are treated ultra-relativistically, helium component was not considered, Saha formula was assumed to be applicable and others. Some of the approximations, evidently, underestimate the effect, but not all. 
A thermodynamic treatment is one of the crucial points here since it gives much greater result than that obtained by consideration of only ionization processes themselves induced by evaporated particles \cite{2008PhRvD..78b3004T, 2009JCAP...10..009C, 2015PAN............}. In fact, in the first case, any energy transferred from the evaporated particles to the baryonic matter contributes to the effect but not only one which is higher than ionization potential as in the second case.



\section {Acknowledgments}


We would like to thank A.~Grobov and S.~Rubin, cooperative work with whom initiated this paper, also E.~Esipova for the help in calculations. The work was supported by grant of RFBR \textnumero 14-22-03048 and in part by grants of RFBR \textnumero 14-22-03031 and of Ministry of Education and Science of the Russian Federation, \textnumero 3.472.2014/K.

\bibliographystyle{utphys}
\bibliography{Article}

\end{document}